\documentclass{sig-alternate-05-2015}
\newfont{\mycrnotice}{ptmr8t at 7pt}
\newfont{\myconfname}{ptmri8t at 7pt}
\let\crnotice\mycrnotice%
\let\confname\myconfname%
\permission{Permission to make digital or hard copies of all or part of this work for personal or classroom use is granted without fee provided that copies are not made or distributed for profit or commercial advantage and that copies bear this notice and the full citation on the first page. Copyrights for components of this work owned by others than ACM must be honored. Abstracting with credit is permitted. To copy otherwise, or republish, to post on servers or to redistribute to lists, requires prior specific permission and/or a fee. Request permissions from Permissions@acm.org.}
\CopyrightYear{2016} 
\setcopyright{acmcopyright}
\conferenceinfo{WNS3,}{June 15-16, 2016, Seattle, WA, USA}
\isbn{978-1-4503-4216-2/16/06}\acmPrice{\$15.00}
\doi{http://dx.doi.org/10.1145/2915371.2915380}

\usepackage{enumitem}
\usepackage{afterpage}
\usepackage[usenames, dvipsnames]{color}
\usepackage{listings}
\usepackage{xcolor}
\usepackage{graphicx}
\usepackage{float}
\usepackage{caption}
\DeclareCaptionType{copyrightbox}
\usepackage{subcaption}
\usepackage{placeins}
\usepackage{url}
\usepackage{epstopdf}
\usepackage{textcomp}

\begin{document}
\title{A Framework for End-to-End Evaluation of 5G mmWave Cellular Networks in ns-3}
\numberofauthors{3} 
%

\author{
%
%
\alignauthor
\makebox[.6\linewidth]{Russell Ford, Menglei Zhang, Sourjya Dutta} 
\makebox[.6\linewidth]{Marco Mezzavilla, Sundeep Rangan}\\
\makebox[.6\linewidth]{\affaddr{New York University}}\\
\makebox[.6\linewidth]{\affaddr{Brooklyn, New York, USA}}\\
\makebox[.6\linewidth]{\email{\{russell.ford,menglei,sdutta\}@nyu.edu}}
\makebox[.6\linewidth]{\email{\{mezzavilla,srangan\}@nyu.edu}}
\and
\makebox[.1\linewidth]~
\and
\makebox[.3\linewidth]{Michele Zorzi}\\
\makebox[.3\linewidth]{\affaddr{University of Padova}}\\
\makebox[.3\linewidth]{\affaddr{Padova, Italy}}\\
\makebox[.3\linewidth]{\email{zorzi@dei.unipd.it}}
}

\maketitle

\begin{abstract}
The growing demand for ubiquitous mobile data services along with the scarcity of spectrum in the sub-6 GHz bands has given rise to the recent interest in developing wireless systems that can exploit the large amount of spectrum available in the millimeter wave (mmWave) frequency range. Due to its potential for multi-gigabit and ultra-low latency links, mmWave technology is expected to play a central role in 5th Generation (5G) cellular networks. Overcoming the poor radio propagation and sensitivity to blockages at higher frequencies presents major challenges, which is why much of the current research is focused at the physical layer. However, innovations will be required at all layers of the protocol stack to effectively utilize the large air link capacity and provide the end-to-end performance required by future networks. 

Discrete-event network simulation will be an invaluable tool for researchers to evaluate novel 5G protocols and systems from an end-to-end perspective. In this work, we present the first-of-its-kind, open-source framework for modeling mmWave cellular networks in the ns-3 simulator. Channel models are provided along with a configurable physical and MAC-layer implementation, which can be interfaced with the higher-layer protocols and core network model from the ns-3 LTE module to simulate end-to-end connectivity. The framework is demonstrated through several example simulations showing the performance of our custom mmWave stack.
\end{abstract}




\vfill

\section{Introduction}
Millimeter Wave (mmWave) communications promise to be highly disruptive for both cellular and wireless LAN technologies due to the potential for multi-gigabit wireless links, which make use of the gigahertz of contiguous bandwidth available at mmWave frequencies in combination with high-dimension antenna arrays for high-gain directional transmission. Although the mmWave channel is known to suffer from poor high-frequency propagation loss, advances in physical layer technology such as adaptive smart antennas, along with recent work on channel measurements and modeling, have paved the way for achieving sufficient range and coverage in these networks \cite{RanRapE:14,AkdenizCapacity:14}. Nevertheless, before mmWave technology can be effectively realized in 5G cellular networks, there are numerous challenges to be addressed, not only at the physical layer, but at higher layers of the radio stack and in the core network as well. For instance, the extreme susceptibility of mmWave links to shadowing from blockages will require frequent, near instantaneous handovers between neighboring cells, fast link adaptation, and a TCP congestion control algorithm that can utilize the large capacity when available but adapt quickly to rapid channel fluctuations to avoid congestion. Therefore, the constraints and characteristics of the mmWave physical layer will require novel solutions throughout the 5G network and across all layers of the stack.
 
Discrete-event network simulators have, for long, been one of the most powerful tools available to researchers for developing new protocols and simulating complex networks. The ns-3 network simulator \cite{ns3Sim} currently implements a wide range of protocols in C++, making it especially useful for cross-layer design and analysis. 

In this work, we present the current state of the millimeter wave module for ns-3, first introduced in \cite{MezzavillaNs3:15}, which can now be easily interfaced with the LTE LENA module \cite{ns3Lena} radio stack and core network in order to evaluate cross-layer and end-to-end performance of 5G mmWave networks. We provide an overview of the module and discuss a number of enhancements and added features since the first version, such as improved statistical channel model derived from 28 GHz channel measurements as well as a new ray tracing-based model. Custom implementations of an ``LTE-like'' Physical (PHY) and Medium Access Control (MAC) layer are also provided, which follow the LENA module architecture. The PHY and MAC classes are parameterized and highly customizable in order to be flexible enough for testing different designs and numerologies without major modifications to the source code.

The rest of the paper is organized as follows. In Section \ref{sec:framework}, we introduce the overall architecture of the mmWave module framework. We then take a closer look at each component, starting with the PHY layer and channel models in Sections \ref{sec:phy}. Section \ref{sec:mac} follows with a discussion on the MAC layer, which includes several scheduler classes as well as support for Adaptive Modulation and Coding (AMC) and Hybrid Automatic Repeat Request (HARQ). In Section \ref{sec:results}, to demonstrate how the framework can be used for cross-layer and end-to-end evaluation, we provide some example simulations showing (i) the capacity of a TDMA mmWave cell with multiple users and (ii) the performance of TCP for a single user under varying channel conditions. Finally, we discuss future work and conclude the paper in Section \ref{sec:conclusions}.

\section{mmWave Framework Overview}
\label{sec:framework}
Presently, the ns-3 mmWave module is targeted for simulating LTE-style cellular networks and is based heavily on the architecture and design patterns of the LTE LENA module. The main enhancement introduced in this latest version of the module is the implementation of Service Access Points (SAPs) for interfacing with the existing LENA classes, which enables mmWave classes to leverage the robust suite of LTE/EPC protocols the LENA module provides. 

\begin{figure}
\includegraphics[width=0.5\textwidth,trim = 1cm 1cm 1cm 1cm,clip] {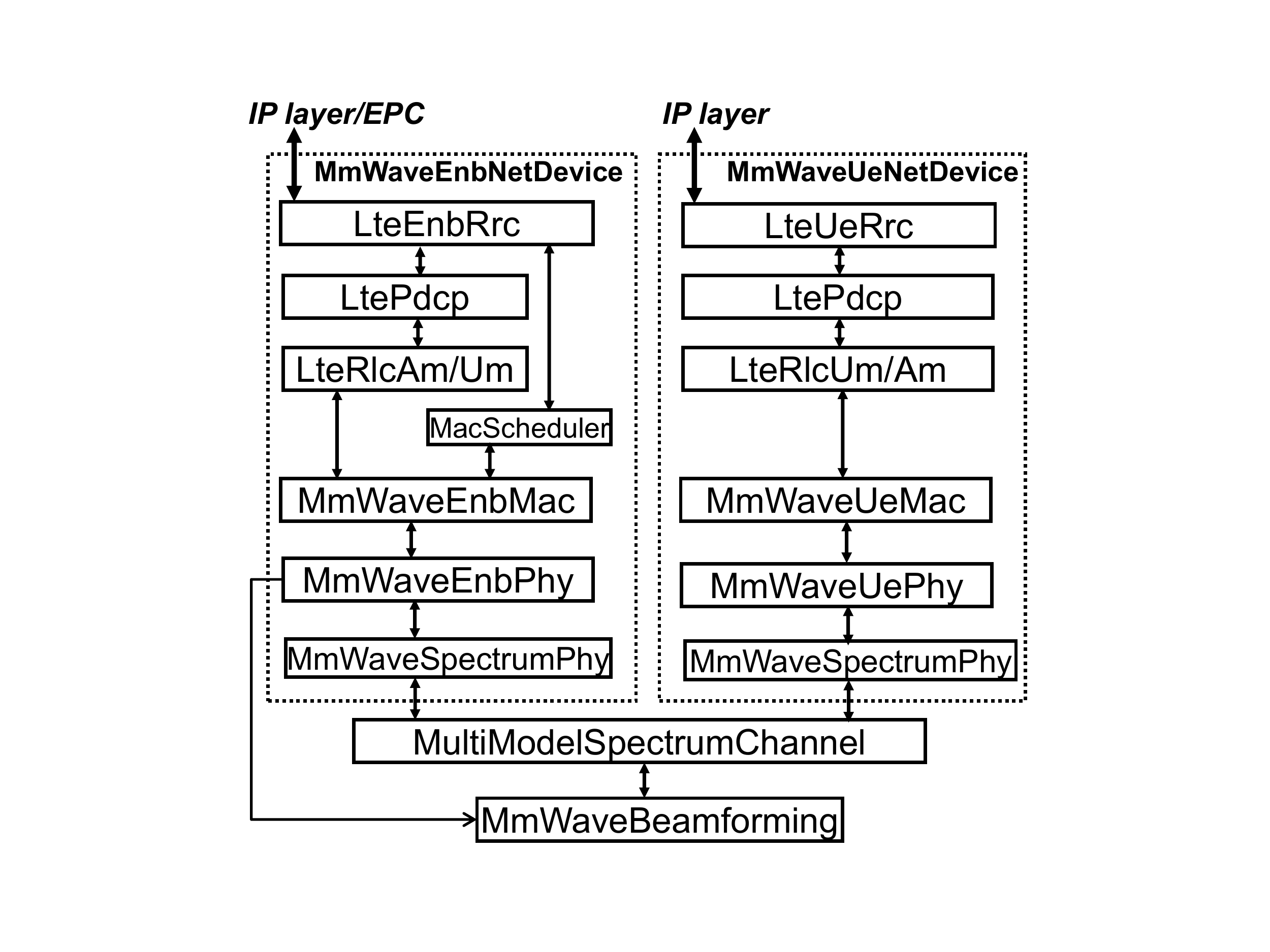}
\caption{\bf Simplified class diagram for the mmWave module.}
\label{fig:class_diagram}
\end{figure}

In Figure \ref{fig:class_diagram}, we show the high-level composition of the \linebreak \texttt{MmWaveEnbNetDevice} and \texttt{MmWaveUeNetDevice} classes, which respectively represent the mmWave eNodeB (eNB) base station and User Equipment (UE) radio stacks. A more detailed UML class diagram is given in Figure 1 of \cite{MezzavillaNs3:15} and details on each layer will be given in their respective sections. The \texttt{MmWaveEnbMac} and \texttt{MmWaveUeMac} MAC layer classes implement the LTE module Service Access Point (SAP) \textit{provider} and \textit{user} interfaces, which allow them to interoperate with the LTE Radio Link Control (RLC) layer. Support for RLC Saturation Mode (SM), Unacknowleged Mode (UM) and Acknowledged Mode (AM) is built into the MAC and scheduler classes (i.e., \texttt{MmWaveMacScheduler} and derived classes). The MAC scheduler also implements a SAP for configuration by the LTE RRC layer (\texttt{LteEnbRrc}). Therefore, all the components required for Evolved Packet Core (EPC) connectivity are available. 

The \texttt{MmWavePhy} classes handle directional transmission and reception of the downlink (DL) and uplink (UL) data and control channels based on control messages from the MAC layer. Similar to the LTE module, each PHY instance communicates over the channel (i.e., \texttt{SpectrumChannel}) via an instance of the \texttt{MmWaveSpectrumPhy} class, which is shared for both DL and UL (instead of separating such objects as in LTE LENA). Instances of \texttt{MmWaveSpectrumPhy} encapsulate all PHY-layer models including those for interference calculation \linebreak (\texttt{MmWaveInterference}), Signal to Interference and Noise Ratio (SINR) calculation (\texttt{MmWaveSinrChunkProcessor}), the Mutual Information (MI)-based error model \linebreak (\texttt{MmWaveMiErrorModel}), which computes packet error probability, as well as the Hybrid ARQ PHY-layer entity \linebreak (\texttt{MmWaveHarqPhy}) for performing soft combining.

A more detailed exposition of the procedures and interactions of these classes is given in \cite{MezzavillaNs3:15}. Since the structure, high-level functions and naming scheme of each class closely follows the LTE LENA module, the reader is also referred to the LENA project documentation \cite{ns3LenaDoc} for more information.

\section{Physical Layer}
\label{sec:phy}
In this section, we discuss the key features of the mmWave PHY layer. Specifically, we have implemented a TDD frame and subframe structure, which has similarities to TD-LTE, but allows for more flexible allocation and placement of control and data channels within the subframe and is suitable for the \textit{variable Transmission Time Interval (TTI)} MAC scheme described in Section \ref{sec:mac}. Significant improvements have also been made to the channel model since the version introduced in \cite{MezzavillaNs3:15}. The 28 GHz statistical path loss model can now be combined with the \textit{building obstacle} model to simulate a realistic shadowing environment. A ray tracing-based path loss and fading model, which makes use of paths generated by third-party ray tracing software, has also been added. Additionally, we have modified the LENA error model and Hybrid ARQ model to be compatible with our custom mmWave PHY and numerology (for instance, to support larger TB and codeword sizes as well as multi-process stop-and-wait HARQ for both DL and UL).
  
%

\subsection{Frame Structure}
\label{subsec3.1}
It is widely agreed that 5G mmWave systems will target Time Division Duplex (TDD) operation because it offers improved utilization of wider bandwidths and the opportunity to take advantage of channel reciprocity for channel estimation \cite{RanRapE:14,PiSysDes:11,AGhosh:14}. In addition, shorter symbol periods and/or slot lengths have been proposed in order to reduce radio link latency \cite{levanen2014radio,Dutta_eucnc:15}. The ns-3 mmWave module therefore implements a TDD frame structure which is designed to be configurable and supports short slots so as to be useful for evaluating different potential designs and numerologies. These parameters, shown in Table {\ref{tab:phy_mac_params}}, are accessible through the common \texttt{MmwavePhyMacCommon} class, which stores all user-defined configuration parameters used by the PHY and MAC classes.

The frame and subframe structures share some similarities with LTE in that each frame is subdivided into a number of subframes of fixed length. However, in this case, the user is allowed to specify the subframe length in multiples of Orthogonal Frequency-Division Multiplexing (OFDM) \linebreak symbols.\footnote{Although many waveforms are being considered for 5G cellular, OFDM is still viewed as a possible candidate. Therefore we adopt OFDM, which allows us to continue to leverage the existing PHY models derived for OFDM from the LTE LENA module.}
Within each subframe, a variable number of symbols can be assigned by the MAC scheduler and designated for either control or data channel transmission. The MAC entity therefore has full control over multiplexing of physical channels within the subframe, as discussed in Section \ref{sec:mac}. Furthermore, each variable-length time-domain data slot can be allocated by the scheduler to different users for either DL or UL.

\begin{figure}[!t]
\includegraphics [width=0.5\textwidth,trim={4.5cm 4cm 3cm 3cm},clip] {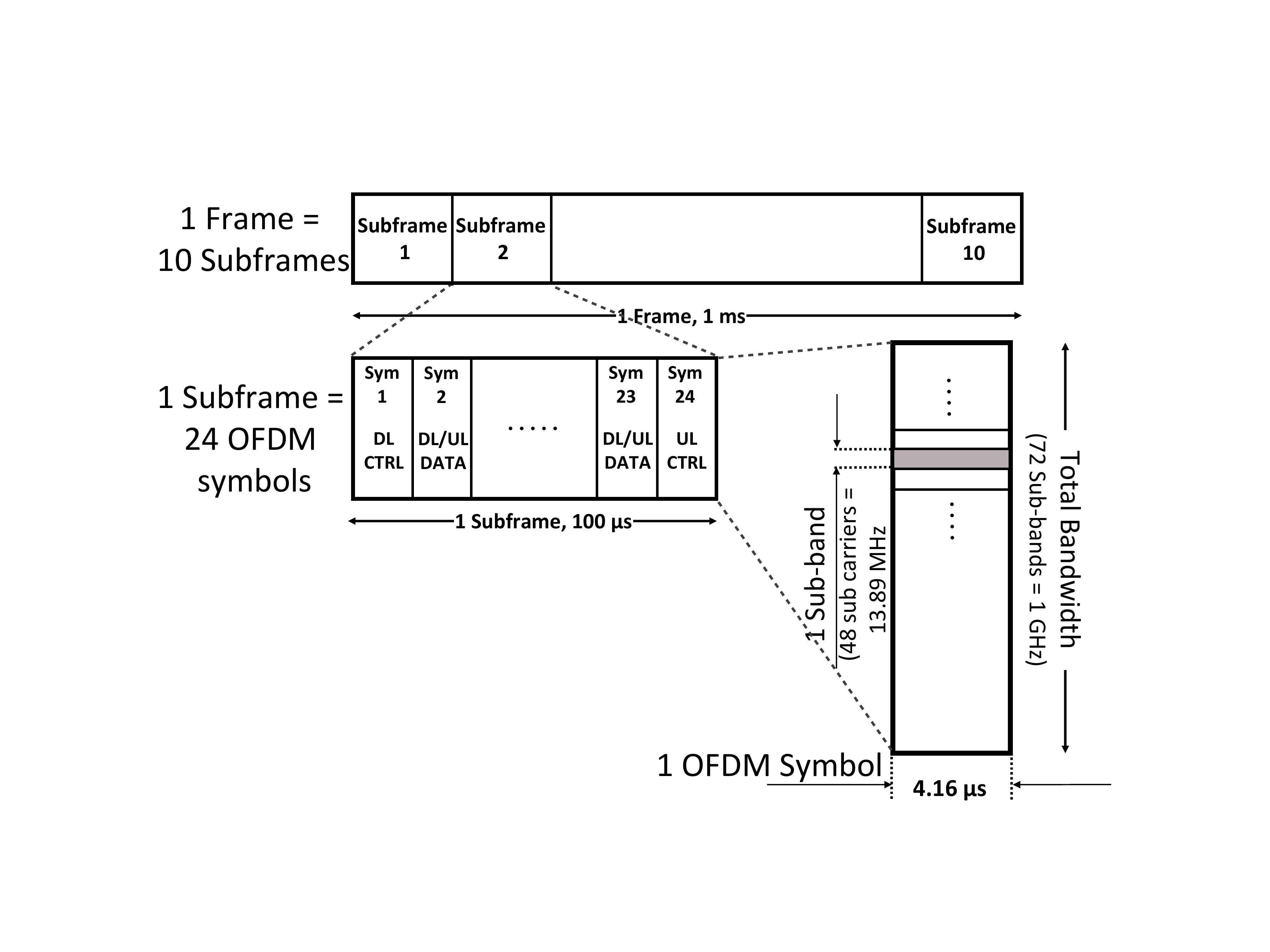}
\caption{\bf Proposed mmWave frame structure.}
\label{fig:frame_structure}
\end{figure}

Figure \ref{fig:frame_structure} shows an example of the frame structure with the numerology taken from our proposed design in \cite{Dutta_eucnc:15}. Each frame of length 1 ms is split in time into 10 subframes, each of duration $100~\mu s$, representing 24 symbols of approximately  $4.16~\mu s$ in length. In this particular scheme, the DL and UL control channels are always fixed in the first and last symbol of the subframe, respectively. A switching guard period of one symbol is introduced every time the direction changes from UL to DL. In the frequency domain, the entire bandwidth of 1 GHz is divided into 72 sub-bands of width 13.89 MHz, each of which is composed of 48 sub-carriers. It is possible to assign UE data to each of these sub-bands, as is done with Orthogonal Frequency-Division Multiple Access (OFDMA) in LTE, however only TDMA operation is currently supported for reasons we shall explain shortly.



\begin{table*}[ht!]
\centering
\caption{\bf Parameters for configuring the mmWave PHY.}
\begin{tabular}{|| l | l | l ||}
\hline \hline
{\bf Parameter Name} & {\bf  Default Value} & {\bf Description} \\
\hline
{\it SubframePerFrame }&  10 & Number of subframes in one frame \\ \hline
{\it SubframeLength} & $100$ & Length of one subframe in \textmu s \\ \hline
{\it SymbolsPerSubframe} & 24 & Number of OFDM symbols per slot \\ \hline
{\it SymbolLength} & $4.16$ & Length of one OFDM symbol in \textmu s \\ \hline
{\it NumSubbands} & 72 & Number of sub-bands \\ \hline
{\it SubbandWidth} & $13.89\times10^6$ & Width of one sub-band in Hz \\ \hline
{\it SubcarriersPerSubband} & 48 & Number of subcarriers in each sub-band \\ \hline
{\it CenterFreq} & $28\times10^9$ & Carrier frequency in Hz \\ \hline
{\it NumRefScPerSymbol }&  864 (25\% of total) & Reference subcarriers per symbol \\ \hline
{\it NumDlCtrlSymbols }&  1 & Downlink control symbols per subframe \\ \hline
{\it NumUlCtrlSymbols }&  1 & Uplink control symbols per subframe \\ \hline
{\it GuardPeriod }&  $4.16$ \textmu s& Guard period for UL-to-DL mode switching \\ \hline
{\it MacPhyDataLatency }&  2 & Subframes between MAC scheduling request and scheduled subframe  \\ \hline
{\it PhyMacDataLatency }&  2 & Subframes between TB reception at PHY and delivery to MAC \\ \hline
{\it NumHarqProcesses }&  20 & Number of HARQ processes for both DL and UL \\ \hline
\hline \hline
\end{tabular}
\label{tab:phy_mac_params}
\end{table*}

%

\subsection{PHY Transmission and Reception}
The \texttt{MmWaveEnbPhy} and the \texttt{MmWaveUePhy} classes model the physical layer for the mmWave eNodeB and the UE, respectively, and encapsulate similar functionality to the \texttt{LtePhy} classes from the LTE module. Broadly, these objects (i) handle the transmission and reception of physical control and data channels (analogous to the PDCCH/PUCCH and PDSCH/PUSCH channels of LTE), (ii) simulate the start and the end of frames, subframes and slots and (iii) deliver received and successfully decoded data and control packets to the MAC layer. 

In \texttt{MmWaveEnbPhy}/\texttt{MmWaveUePhy}, calls to \texttt{StartSubFrame()} and \texttt{EndSubFrame()} are scheduled at fixed periods, based on the user-specified subframe length, to mark the start and the end of each subframe. The timing of variable-TTI slots, controlled by scheduling the \texttt{StartSlot()} and \texttt{EndSlot()} methods, is dynamically configured by the MAC via the MAC-PHY SAP method \texttt{SetSfAllocInfo()}, which enqueues a \texttt{SfAllocInfo} allocation element for some future subframe index specified by the MAC. A \textit{subframe indication} to the MAC layer triggers the scheduler at the beginning of each subframe to allocate a future subframe. For the UE PHY, \texttt{SfAllocInfo} objects are populated after reception of Downlink Control Information (DCI) messages. At the beginning of each subframe, the current subframe allocation scheme is dequeued, which contains a variable number of \texttt{SlotAllocInfo} objects. These, in turn, specify contiguous ranges of OFDM symbol indices occupied by a given slot, along with the designation as either \textit{DL} or \textit{UL} and control (\textit{CTRL}) or data (\textit{DATA}). 

The data packets and the control messages generated by the MAC are mapped to a specific subframe and slot index in the \textit{packet burst map} and \textit{control message map}, respectively. Presently, in our custom subframe design, certain control messages which must be decoded by all UEs (such as the DCIs) are always transmitted in fixed PDCCH/PUCCH symbols as the first and last symbol of the subframe, but this static mapping can easily be changed by the user. \footnote{As in \cite{levanen2014radio,Dutta_eucnc:15}, we assume that either FDMA or SDMA-based multiple access would be used in the control regions. However, we do not currently model these modulation schemes nor the specific control channel resource mapping explicitly. We intend for this capability to be available in later versions, which will enable more accurate simulation of the control overhead.} 
Other UE-specific control and data packets are dequeued at the beginning of each allocated TDMA data slot and are transmitted to the intended device.

To initiate transmission of a data slot, the eNB PHY first calls \texttt{AntennaArrayModel::ChangeBeamformingVector()} to update the transmit and receive beamforming vectors for both the eNB and the UE. In the case of control slots, no beamforming update is applied since we currently assume an ``ideal'' control channel. For both DL and UL, either the \linebreak \texttt{MmWaveSpectrumPhy} method \texttt{StartTxDataFrame()} or \linebreak \texttt{StartTxCtrlFrame()} is then called to transmit a data or control slot, respectively. The functions of \texttt{MmWaveSpectrumPhy}, which are similar to the corresponding LENA class, are as follows. After the reception of data packets, the PHY layer calculates the SINR of the received signal in each sub-band, taking into account the path loss, MIMO beamforming gains and frequency-selective fading. This triggers the generation of Channel Quality Indicator (CQI) reports, which are fed back to the base station in either UL data or control slots. The error model instance is also called to probabilistically compute whether a packet should be dropped by the receiver based on the SINR and, in the case of a HARQ retransmission, any soft bits that have been accumulated in the PHY HARQ entity (see Section \ref{sec:harq}). Uncorrupted packets are then received by the \texttt{MmWavePhy} instance, which forwards them up to the MAC layer SAP.

\subsection{Channel Models}
\label{sec:channel_model}
The mmWave module allows the user to choose between two channel models. The first, implemented in the \texttt{MmWavePropagationLossModel} and \texttt{MmWaveBeamforming} classes, is based on our previous code in \cite{MezzavillaNs3:15}, which is derived from extensive $\text{\textsc MATLAB}^{\text{\textregistered}}$ simulation of the 28 GHz channel presented in \cite{AkdenizCapacity:14}. This model has now been combined with the built-in \texttt{BuildingsObstaclePropagationLossModel} for simulating mobility. The second model, the \texttt{MmWaveChannelRaytracing} class, uses data obtained from  third-party ray tracing software. The details of each model are as follows.


\subsubsection{Simulation-Generated Statistical Model}
\label{sec:statistical_model}
A full discussion on the simulation based channel model is available in \cite{MezzavillaNs3:15}. We have strengthened this model by incorporating a building obstacle propagation model which enables simulation of UE mobility through a shadowing environment. For each simulation, instances of the building class (built into ns-3) are used to simulate obstacles. The channel state is updated based on the relative position of the transmitter, the receiver and the buildings as the UE moves through the environment. A virtual line is drawn between the transmitter and the receiver. If this line intersects any building we assign the channel state as NLoS; otherwise, we assign a LoS channel.



After selecting the channel state, the propagation loss can be computed as
\begin{equation}
PL(d)[dB] = \alpha+\beta10 \log_{10}(d)+\xi, \quad
\xi \sim N(0,\sigma^{2}),
\label{pathloss}
\end{equation}
where $\xi$ represents shadowing, $d$ is the distance from receiver to transmitter and the values of the parameters $\alpha$, $\beta$, and $\sigma$ for each channel state are given in \cite{AkdenizCapacity:14}.
In this model, we consider the channel to be in outage if the distance between the transmitter and the receiver exceeds a predefined threshold.

Due to lack of information about the effects of diffraction in the mmWave channel, we assume the propagation loss increases or decreases suddenly without a transition period when the channel state changes. We plan to model diffraction more accurately in our future work.

\paragraph*{Channel matrix}

Following the method in \cite{AkdenizCapacity:14}, to compute the long-term statistical characterization of the mmWave channel, we model it as a combination of clusters, each composed of several subpaths. The channel matrix is described by the following equation,
\begin{equation}
\label{eqn:channel_gain}
H(t,f)=\sum_{k=1}^{K}\sum_{l=1}^{L_k}g_{kl}(t,f){\bf u}_{rx}(\theta^{rx}_{kl},\phi^{rx}_{kl}){\bf u}^*_{tx}(\theta^{tx}_{kl},\phi^{tx}_{kl})
\end{equation}
where {$K$} is the number of clusters, {$L_k$} the number of subpaths in cluster $k$, {$g_{kl}(t,f)$} the small-scale fading over frequency and time, {$\bf{u}_{rx}(\cdot)$}is the spatial signature of the receiver and {$\bf{u}_{tx}(\cdot)$} the spatial signature of the transmitter.

As given in \cite{AkdenizCapacity:14}, the small-scale fading is generated by
\begin{equation}
g_{kl}(t,f)=\sqrt{P_{kl}}e^{2\pi if_{d}cos(\omega_{kl})t-2\pi i\tau _{kl}f}
\label{eq:smallscale}
\end{equation}
where {$P_{kl}$} is the power of subpath $kl$, {$f_{d}$} is the maximum Doppler shift, {$\omega_{kl}$} is the angle between subpath $kl$ and the direction of motion of the receiver, {$\tau _{kl}$} gives the delay spread and {$f$} is the carrier frequency.

\paragraph*{Beamforming} 
The two methods implemented in the \linebreak \texttt{MmWaveBeamforming} class to compute beamforming vectors are the \textit{power iteration method} and the \textit{swipe sector method}. For the power iteration method, we assume that the BS knows the channel matrices and can compute the largest singular value and singular vector associated with the strongest path (i.e., it uses non-codebook based beamforming with perfect Channel State Information (CSI)). Therefore, the optimal set of TX/RX beamforming vectors will always be selected to maximize the antenna array gain for transmission between a given BS-UE pair. 

The swipe sector method implements a basic cell search/ synchronization technique where the cell is divided into fixed sectors. The BS and UE scan all sectors and select the beam with maximum gain based on these pre-stored beamforming vectors  (i.e., they perform codebook-based beam switching). This method does not require the CSI to be known a priori, but takes additional time to scan the cell. We provide this code as a basis for implementing more advanced cell search and synchronization protocols.


\paragraph*{Channel Configuration} 
For both methods, the channel matrices and optimal beamforming vectors are pre-generated in $\text{\textsc MATLAB}^{\text{\textregistered}}$ to reduce the computational overhead in ns-3. At the beginning of each simulation we load 100 instances of the spatial signature matrices, along with the beamforming vectors.

In order to simulate realistic channels with large-scale fading, the channel matrices are updated periodically for NLoS channels but remain constant for LoS links as they are inherently more stable. Currently, no results are available for how the large-scale statistics of the mmWave channel change over time for a mobile user. We therefore implement a form of large-scale \textit{block fading}, where we update the channel by randomly selecting one of the 100 channel matrix-beamforming vector combinations after some interval. The large-scale parameters of the channel are thus independent in each interval. The update time can be some fixed interval specified by the \texttt{LongTermUpdatePeriod} attribute of the \texttt{MmWaveBeamforming} class. We also provide the option to update after some time drawn from an exponentially-distributed random variable (i.e., a Poisson process) with mean also defined by the update period attribute. It should be noted that the accuracy of this method is not validated at this time.
  
The small-scale fading is calculated for every transmission based on Equation (\ref{eq:smallscale}) where we obtain the speed of the user directly from the mobility model. The remaining parameters depending on the environment are assumed to be constant over the entire simulation time. 

\begin{figure}[t!]
    \centering
    \includegraphics[ width=0.5\textwidth]{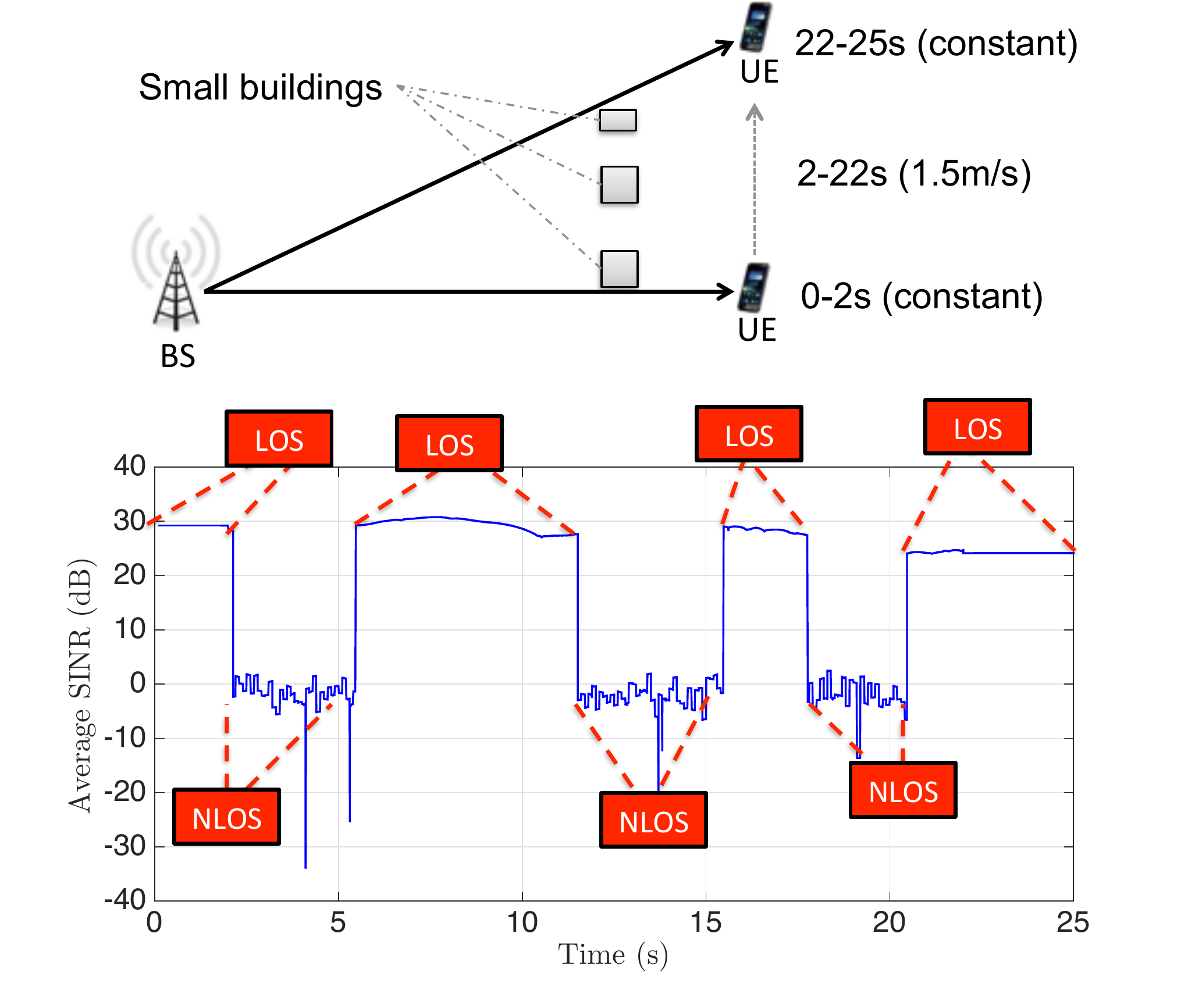}
    \caption{\textbf{Average SINR plot for simulated route.}}
    \label{fig:simSINR}
\end{figure}

From Figure \ref{fig:simSINR} we observe the average SINR trend obtained in a scenario where 3 buildings are distributed between BS and UE. The number of antennas at the BS and the UE is 64 and 16, respectively. The user starts moving at a speed of 1.5 m/s 2 seconds after the start of the simulation and stops after 20 seconds. As expected, the SINR is constant over time while the user is static (0-2 s and 22-25 s). However, the SINR varies over time when the user is in motion. The sudden SINR jumps result from the switching of the channel state (as discussed, the channel matrices are updated after a fixed 100 ms interval for the NLoS channel and remain unchanged for LoS transmissions).

\subsubsection{Ray Tracing-Generated Model}
\begin{figure}[t!]
    \centering
    \includegraphics[width=0.45\textwidth]{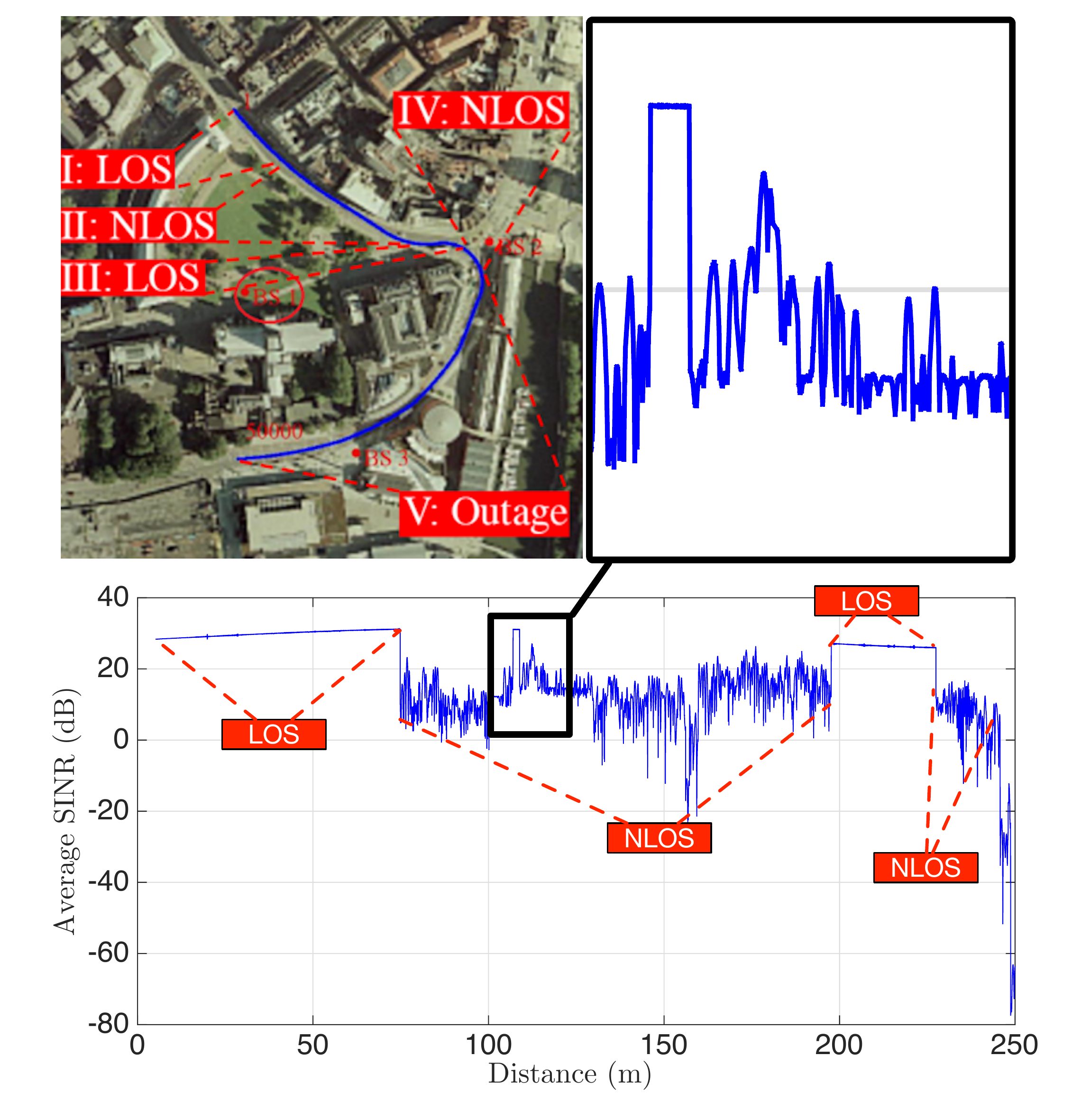}
    \caption{\textbf{Average SINR plot for raytracing route.}}
    \label{fig:raytra1}
\end{figure}

In order to better characterize the time dynamics of the \linebreak mmWave channel, we have added a ray tracing-generated channel model that computes the channel matrices in ns-3. The input is traces generated by third-party ray-tracing software, which simulates the physics of radio propagation for a specific environment. 
\footnote{The ray tracing data is provided by the Communication Systems and Networks Group, University of Bristol, UK.} 
This channel model offers more flexibility to customize parameters, such as transmit and receiver antenna elements. Figure \ref{fig:raytra1} shows the average SINR generated in ns-3 over a given ray tracing route, with 64 transmit and 16 receive antennas. Propagation loss and channel matrices are computed according to Equation (\ref{eqn:channel_gain}), where parameters are obtained from the ray-tracing data containing 5000 samples within a 500 meter-long route. Each sample includes the following fields: 
\begin{itemize}[noitemsep]
  \item Number of paths
  \item Propagation loss per path
  \item Delay per path
  \item Angle of arrival (elevation and azimuth plane) per path
  \item Angle of departure (elevation and azimuth plane) per path
\end{itemize}

As the user moves, the channel matrices are updated accordingly. For example, if the current location of the user is 10.1 meters from the BS, channel matrices are computed using the data corresponding to this distance. The beamforming vectors are generated using the power method discussed earlier.

Figure \ref{fig:raytra1} plots the average SINR indicating both LoS intervals and NLoS channel states. The SINR has a sudden change when the channel state switches. We note that the SINR curve within LoS matches our simulation generated model, but for NLoS the ray tracing based model introduces more random variation.

\subsubsection{Interference}
\label{sec:interference}
Albeit potentially less significant for directional mmWave signals, which are generally assumed to be power-limited, there are still some cases where interference is non-negligible. For instance, although intra-cell interference (i.e., from devices of the same cell) can be neglected in TDMA or FDMA-mode operation, it does need to be explicitly calculated in the case of SDMA/Multi-User MIMO, where users are multiplexed in the spatial dimension but operate in the same time-frequency resources. As mentioned, only TDMA transmission is currently supported, however we anticipate that other multiple access schemes, including SDMA, will be added in later versions. Therefore, we include interference computation in the \texttt{MmWaveInterference} class that takes into account the beamforming vectors associated with each link. More details on the relevant computation can be found in Section 3.4 of \cite{MezzavillaNs3:15}.

\section{MAC Layer}
\label{sec:mac}
The high-level structure and functions of the mmWave MAC layer are now introduced. In particular, the two scheduler classes, multi-process stop-and-wait Hybrid ARQ (for both the DL and the UL), along with some minor modifications that have been made to the LENA module AMC and MI error model classes, are described in the following sections.

%


\subsection {MAC Scheduling}
\label{sec:scheduler}
We now present the implementation of two scheduler classes, which are based on a variable-length or \textit{flexible} TTI, Time-Division Multiple Access (TDMA) scheme. 

TDMA is widely assumed to be the de-facto scheme for mmWave access because of the dependence on analog beamforming, where the transmitter and receiver align their antenna arrays to maximize the gain in a specific direction (rather than in a wide angular spread or omni-directionally, like conventional antennas). Many early designs and prototypes have been TDMA-based \cite{PiSysDes:11,AGhosh:14}, with others incorporating SDMA for the control channel only \cite{levanen2014radio}. While SDMA or FDMA schemes (like in LTE) are possible with \textit{digital beamforming}, which would allow the base station to
transmit and receive in multiple directions within the same time slot, they may not be practical for mmWave systems due to high power consumption from requiring Analog-to-Digital Converters (ADCs) on each antenna element \cite{Dutta_eucnc:15}. It should also be noted that FDMA with analog beamforming is possible with wider beam widths, but this approach comes at the cost of some of the directional gain.

Furthermore, one of the foremost considerations driving innovation for the 5G MAC layer is latency. The Key Performance Indicator (KPI) of 1 ms over-the-air latency has been proposed as one of the core 5G requirements by such standards bodies as the ITU as well as recent studies such as those carried out under the METIS 2020 project \cite{popovsk2013eu}. However, a well-known drawback of TDMA is that fixed slot lengths or TTIs can result in poor resource utilization and latency, which can become particularly severe in scenarios where many intermittent, small packets must be transmitted to/received from many devices. 

Based on these considerations, variable TTI-based TDMA frame structures and MAC schemes have been proposed in \cite{levanen2014radio, Dutta_eucnc:15, kela2015novel}. This approach supports slot sizes that can vary according to the length of the packet or Transport Block (TB) to be transmitted and are well-suited for diverse traffic since they allow both bursty or intermittent traffic with small packets as well as high-throughput data like streaming and file transfers to be scheduled efficiently and without significant under-utilization. 

\subsubsection{Round-Robin Scheduler}
The \texttt{MmWaveFlexTtiMacScheduler} class is the default scheduler for the mmWave module. It implements a variable TTI scheme previously described in Section \ref{sec:phy} and assigns OFDM symbols to user flows in Round-Robin (RR) order. Upon being triggered by a subframe indication, any HARQ retransmissions are automatically scheduled using whatever OFDM symbols are available. While the slot allocated for a retransmission does not need to start at the same symbol index as the previous transmission of the same TB, they do need the same number of contiguous symbols and Modulation and Coding Scheme (MCS), since an adaptive HARQ scheme has not yet been implemented.

Before scheduling new data, Buffer Status Report (BSR) and Channel Quality Indicator (CQI) messages are first processed. The MCS is then computed by the AMC model for each user based on the CQIs for the DL or SINR measurements for the UL data channel. The MCS and the buffer length of each user are used to compute the minimum number of symbols required to schedule the data in the user's RLC buffers. 

To assign symbols to users, the total number of users with active flows is calculated. Then the total available data symbols in the subframe are divided evenly among users. If a user requires fewer symbols to transmit its entire buffer, then the remaining symbols (i.e., the difference between the available and required slot length) are distributed among the other active users.

One also has the option to set a fixed number of symbols per slot by enabling \textit{fixed TTI} mode. However, utilization and latency are likely to suffer in this case, depending on the traffic pattern.

\subsubsection{Max-Weight Scheduler}
The \texttt{MmWaveFlexTtiMaxWeightMacScheduler} class is similar to the RR scheduler but is intended to provide various priority queue policies. Currently only an Earliest Deadline First (EDF) policy is implemented, which weighs flows by their relative deadlines for packet delivery, with weights determined by the delay budget of the QoS Class Indicator (QCI) configured by the RRC layer. The EDF scheduler can be used to evaluate the delay performance of various radio frame configurations, although the results of such analysis are outside the scope of this paper. Other weight-based disciplines, such as Proportional Fair (PF) scheduling, will be added in future versions.



 
\subsection{Adaptive Modulation and Coding}
\label{sec:amc}

The \texttt{MmWaveAmc} class reuses most of the code from the corresponding LENA module class. Some minor modifications and additional methods were necessary to accommodate the dynamic TDMA MAC scheme and frame structure. For instance, the \texttt{GetTbSizeFromMcsSymbols()} and \texttt{GetNumSymbolsFromTbsMcs()} methods are used by the scheduler to compute the TB size from the number of symbols for a given MCS value, and vice-versa. Also the \texttt{CreateCqiFeedbackWbTdma()} method is added to generate wideband CQI reports for variable-TTI slots. 

\begin{figure}[!t]
\centering
\includegraphics [width=0.43\textwidth,trim=0cm 6cm 0.5cm 6cm,clip] {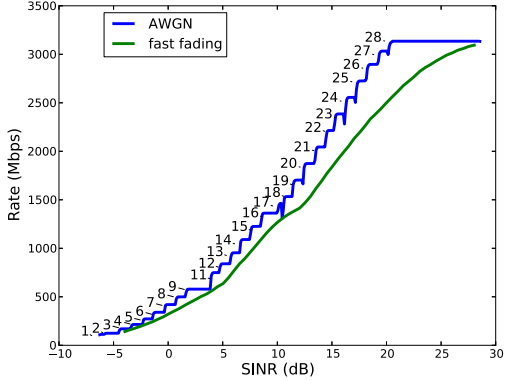}
\caption{\bf Rate and MCS vs. SINR for single user under AWGN and fast-fading mmWave channels}
\label{fig:amc_sinr_rate}
\end{figure}

Figure \ref{fig:amc_sinr_rate} shows the results of the test case provided in \textit{mmwave-amc-test.cc}. This simulation serves to demonstrate the performance of the AMC and CQI feedback mechanisms for a single user in the UL (although a multi-user scenario could easily be configured as well). The default PHY/MAC parameters in Table \ref{tab:phy_mac_params} are used along with the default scheduler and default parameters for the statistical path loss, fading and beamforming models (i.e., \texttt{MmWavePropagationLossModel} and \texttt{MmWaveBeamforming}).

We compute the rate versus the average SINR over a period of 12 seconds (long enough for the small-scale fading to average out), after which we artificially increase the path loss while keeping the UE position fixed. The average PHY-layer rate is then computed as the average sum size of successfully decoded TBs per second. As the SINR decreases, the MAC will select a lower MCS level to encode the data. The test is performed for the AWGN case as well as for small-scale fading. Although the UE position relative to the base station is constant, we can generate time-varying multi-path fading through the \texttt{MmWaveBeamforming} class by setting a fixed speed of 1.5 m/s to artificially generate Doppler, which is a standard technique for such analysis. Also we assume that the long-term channel parameters do not change for the duration of the simulation.

If this plot is compared to the one generated from a similar test in Figure 3.1 of the LENA documentation \cite{ns3LenaDoc}, we notice that the AWGN curve from the mmWave test is shifted by approximately 5 dB to the left, indicating that the LENA version is transitioning to a lower MCS at a much higher SINR. This is because the LENA test is using the more conservative average SINR-based CQI mapping, whereas we use the Mutual Information-Based Effective SINR (MIESM) scheme with a target maximum TB error of 10\% in order to maximize the rate for a given SINR \cite{Mezzavilla:12}. 

\subsection{Hybrid ARQ Retransmission}
\label{sec:harq}
Full support for HARQ with soft combining is now included in the mmWave module. The \texttt{MmWaveHarqPhy} class along with the functionality within the scheduler are based heavily on the LENA module code. However, multiple HARQ processes per user in the UL are now possible. The number of processes can also be configured through the \texttt{NumHarqProcesses} attribute in \texttt{MmWavePhyMacCommon}. Additional modifications were needed to support larger codeword sizes in both the HARQ PHY methods and the error model.

\section{Example Simulations}
\label{sec:results}
We now present some example simulations to show the utility of the framework for the analysis of novel mmWave protocols and testing higher-layer network protocols, such as TCP, over 5G mmWave networks. The simulations in this section are all configured with basic PHY and MAC parameters as in Table \ref{tab:phy_mac_params}, with other notable parameters given in the sequel. 
\vfill
\subsection {Multi-User Throughput Simulation}
\begin{figure}[!t]
\centering
\includegraphics [width=0.42\textwidth] {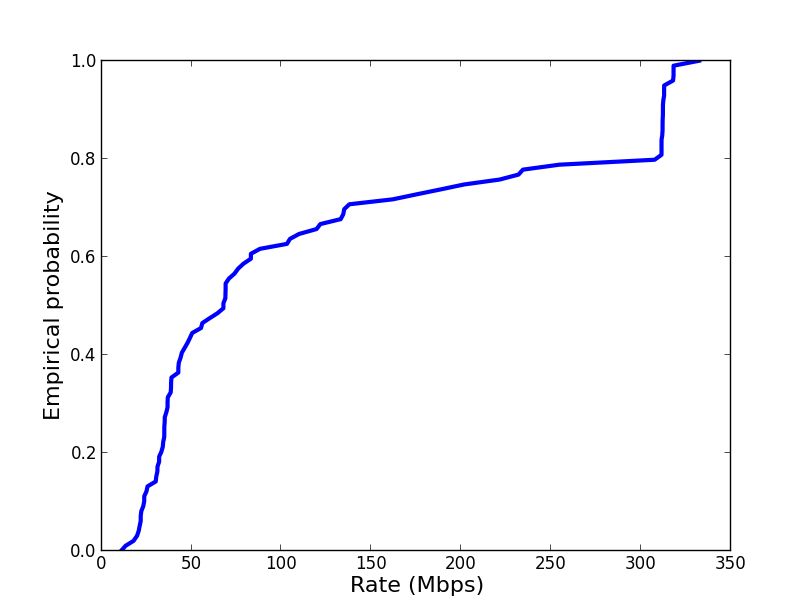}
\caption{\bf Empirical CDF of DL user rates for 10 users with RR scheduling. }
\label{fig:cap_rate_cdf}
\end{figure}
The purpose of this experiment, which one can reproduce by running the \textit{mmwave-tdma} example, is to simulate the DL throughput of 10 UEs in a 1 GHz mmWave cell under the variable TTI/TDMA MAC scheme and round-robin scheduling policy. UEs are placed at uniformly random distances between 20 and 200 meters from a single eNodeB. As explained in Section \ref{sec:amc}, users are stationary but are modeled as having a constant speed of 1.5 m/s and are thus subject to small-scale fading. The long-term channel parameters are updated based on the exponentially-distributed update time with a mean of 100 ms (see Section \ref{sec:channel_model}). Rates are computed from the average size of RLC PDUs delivered to each UE and therefore reflect the performance of the stack up to and including the RLC layer. We assume full-buffer traffic.

The simulation is performed for 10 runs or \textit{drops} of the 10 UEs, where for each drop they are placed at different distances and assigned different channel matrices. The average system throughput at the RLC layer for this scenario over all drops is found to be about 1.2 Gbps. We observe in the plot of the empirical distribution function in Figure \ref{fig:cap_rate_cdf} how UEs with LoS links all have roughly the same average rates around 325 Mbps. Also a significant number of NLoS users achieve rates over 100 Mbps, and even the worst 5\% of users at the cell edge can get between 10 and 20 Mbps. 

\subsection {TCP Performance over mmWave}

Here we run the \textit{mmwave-tcp-building} example to analyze the performance of TCP flows over a mmWave link. TCP data packets are sent from a remote host to the UE at a rate of 1 Gbps. The New Reno algorithm is used for this experiment. The delays for the point-to-point link between the remote host and PDN-Gateway (PGW), as well as that from the PGW to the BS, are set to 10 ms. Thus, the contribution to the total Round Trip Time (RTT) from the core network is 40 ms and any additional latency is due to the radio link, which, under stable queue conditions, is observed to be less than 10 ms. The size of the RLC-AM buffer is adjusted to 10 Megabytes to avoid overflow. The TCP buffer size is set to 5 Megabytes and the slow start threshold is 6000 segments (about 3 MB).

\begin{figure}[t!]
    \centering
    \includegraphics[trim={20mm 1mm 30mm 1mm}, clip, width=0.9\columnwidth]{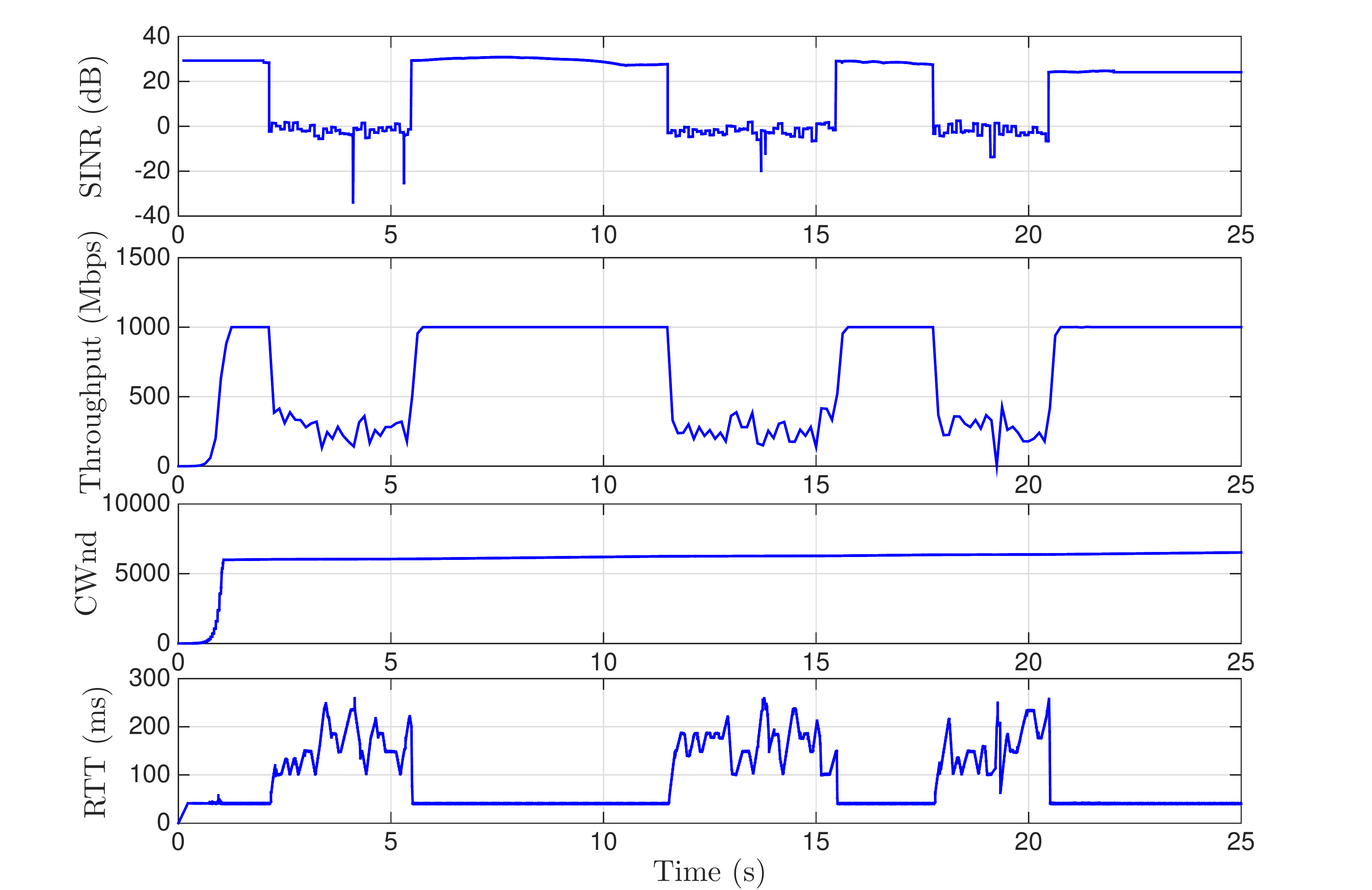}
    \caption{\textbf{TCP performance for simulated route.}}
    \label{fig:s2}
\end{figure}

Figure \ref{fig:s2} plots the SINR, transport layer throughput, congestion window size (CWnd) and  RTT. As shown, the transport layer throughput matches the sending rate for the LoS channel, but is reduced when the channel is in the NLoS state. This is attributed to the MAC-layer AMC model adapting to the change in capacity. From the congestion window plot we see that there is no TCP timeout or retransmission since the packet loss events are handled by lower layer retransmissions, i.e. MAC layer HARQ and RLC ARQ. Moreover, we see that RTT is around the baseline of 40 ms for the LoS channel, but goes above 150 ms for the NLoS case. It is clear that decreased channel capacity and more frequent RLC retransmission events cause the RLC buffer to become backlogged, which explains the increase in RTT. This result suggests the need for a more advanced congestion control mechanism, perhaps aided by feedback or control from lower layers, to prevent large spikes in latency under rapid channel fluctuations.

%

\section{Conclusions \& Future work}
\label{sec:conclusions}
In this paper, the current state of the ns-3 framework for simulation of mmWave cellular systems has been presented. The code, which is publicly available at GitHub \cite{github-ns3-mmwave}, is highly modular and customizable to facilitate researchers to experiment with novel 5G protocols. The code includes implementations of a mmWave eNodeB and User Equipment stack, including the MAC layer, PHY layer and channel models. Some example simulations have been given, which show how the framework may be used for analysis of custom mmWave PHY/MAC protocols as well as higher-layer network protocols over a mmWave stack and channel. 

As part of our future work, we have targeted several new features for channel modeling, including a more accurate model for large-scale fading for mobile users as well as channel matrix generation and beamforming computation within ns-3 to support experimentation with adaptive beamforming algorithms. Future enhancements to the MAC layer include support for other multiple access schemes, relay devices, and additional scheduling algorithms. Also, although a number of configurable example scripts are currently included, which may be used for testing, a complete test framework is not yet been provided. Thus, we intend to include a set of test scripts in a later release.

\FloatBarrier

\balancecolumns
\end{document}